\documentclass[a4paper,12pt]{article}
\usepackage{amsmath}
\usepackage{ulem}
\usepackage{calc}
\usepackage{vmargin}
\usepackage{amstext}
\usepackage{graphicx}
\usepackage{amsfonts}
\usepackage{amssymb}
\usepackage{misccorr}

\begin{document}
	
	\title{		{\bf  Contact mappings of differential equations. } }
	
		\author {Kaptsov O.V.}
	\date{}
	\maketitle

In this paper we consider mappings of jet spaces that preserve the module of canonical  Pfaffian forms, but are not generally invertible. These mappings are called contact. A lemma on the prolongation of contact mappings is proved.	
Conditions are found for which mappings transform solutions of some partial differential equations into ones of other equations. 
Examples of contact mappings of differential equations are given.
%We find conditions that guarantee that contact mappings transform solutions of some partial derivative equations into solutions of other equations.
We consider contact mappings depending on a parameter and give example 
of differential equation invariant under such maps.

 %Contact mappings depending on the parameter are also considered.
 %Examples of contact mappings that transform
 %solutions of some differential equations into solutions of others.
 
	
\section{Introduction}
   
    As is well known, contact transformations are used to solve problems of classical mechanics and equations of mathematical physics
    \cite{Ovs,Arnold,Ibr}.
   Popular examples of such transformations are the Legendre and Ampere transformations. 
   The theory of contact transformations was developed by S. Lie. 
       At present, there is a large literature devoted to these issues \cite{Bluman,Ibr2,Olver,Vinogradov}.
   The contact transformations are diffeomorphisms of the jet space that preserve the contact structure.
   To integrate differential equations, it is useful to find contact transformations that leave these equations invariant.
        
    However, not only contact transformations are applied to integrate differential equations.
   Leonhard Euler started using differential substitutions, which are not diffeomorphisms, to integrate linear partial differential equations \cite{Euler}.
   Now these substitutions are called the Euler-Darboux transformation \cite{Kaptsov} or simply the Darboux transformation \cite{Matveev}.   
      
    In this paper, we consider analytic mappings of jet spaces that preserve the modulus of canonical differential forms and call the
    mappings contact.     
    We prove a lifting lemma that shows how to construct a contact mapping.
            For applications  to differential equations,
    the mappings is required to transform solutions of the equations into solutions of other equations or act on solutions of given equations.
    Examples of second-order partial differential equations connected by contact mappings are given.
    
    We also study contact mappings depending on a parameter.
    It's easier to look for such mappings in the form of series in powers of the parameter.
    As an example, consider the Burgers equation.
    Parametric contact mappings are found that act on solutions of this equation.     These mappings have no inversional maps.

 \section{Contact mappings of the jet space}
 We begin with notations and definitions.
 Denote by $\mathbb{Z}_{\geq 0} $ the non-negative integer numbers and by
  $\mathbb{N}_n$ the set of natural  numbers $1,...,n$.
 The $p$-th-order jet space \cite{Olver} with  coordinates
$\{x_i,u^j_{\alpha}: i\in \mathbb{N}_n,  j\in \mathbb{N}_m, 
\alpha\in\mathbb{Z}_{\geq 0}^n \}$  is denoted by $J^p(\mathbb{R}^n,\mathbb{R}^m)$
 or simply $J^p$. We suppose that $J^0= \mathbb{R}^n(x)\times\mathbb{R}^m(u)$.

 Denote by $J^{\infty}$ the space of infinite jets   and 
 by $\pi_p$ the projection  $\pi_p: J^{\infty} \longrightarrow J^p$.
  % If $a\in J^{\infty}$ we write $a_p=\pi_p(a)$.
Consider a point $a\in J^{\infty}$ and a ring $A^p_a$ of convergent power series centered at the point $a_p=\pi_p(a)$.
We write
   $$ A_a = \bigcup_{p=0}^{\infty} A^p_a .$$

    Recall that an operator $D$ on a ring is called a derivation operator 
   if it satisfies the conditions:    
 $$D (a + b) = Da + Db ,  \qquad   D(ab) = D(a)b+aD(b)$$
  for all elements $a, b$ of the ring.
 We say that a derivation operator $D_k$ $(k\in\mathbb{N}_n)$ on the ring $A_a$ is  the total derivative when the  following conditions is satisfied
   $$  D_k(x_i)= \delta_{ik}, \qquad  D_k( u^j_{\alpha})=u^j_{\alpha+1_k} ,   $$
 where  $\delta_{ik} $ is the Kronecker delta and $1_k=(\delta_{1k},\dots,\delta_{nk})\in\mathbb{Z}_{\geq 0}^n$. 
 So $A_a$ is a differential ring with set $\Delta$ of derivation operators $D_1,\dots,D_n$;
its elements are called differential power series.
 Any ideal of $A_a$ stable under $\Delta$ is called a differential ideal  of $A_a$. 
The differential ideal of the ring $A_a$ generated by the set $E\subset A_a$ is denoted by $<E>$.

The set of differentials
 $$\{dx_i, du^j_{\alpha}: i\in\mathbb{N}_n, j\in\mathbb{N}_m; \alpha\in\mathbb{Z}_{\geq 0}^n, |\alpha|< p \}$$
 generates a left module $\Omega_{a_p}$ of differential 1-forms over the ring $A_{a_p}$.
As usual, we say that differential forms 
\begin{equation} \label{omega}
	\omega^j_{\alpha} = du^j_{\alpha} - \sum_{i=1}^{n} D_i(u^j_{\alpha}) dx_i , \qquad
	j\in\mathbb{N}_m,\quad  \alpha\in\mathbb{Z}_{\geq 0}^n    
\end{equation}
are canonical.

\noindent
{\bf Definition.} A submodule of the left module $\Omega_{a_p}$  generated by canonical forms $\omega^j_{\alpha}$
(where $|\alpha|\leq p$) is denoted by $\mathcal{C}^p_a$ and is called the contact submodule.

We describe below a dual transformation of forms \cite{Munkres}.
Let $\mathcal{A}(W )$ be the ring of analytic functions on an open set $W\subset\mathbb{R}^k$  and let $\Omega (W)$ be the left module of differential 1-forms on $W$. 
Suppose $W_1\subset\mathbb{R}^{k_1}$, $W_2\subset\mathbb{R}^{k_2}$ are open sets. 
Then  any analytic mapping $\phi: W_1 \longrightarrow W_2 $ 
induces a homomorphism
$$\phi^{\star} :  \mathcal{A}(W_2) \rightarrow   \mathcal{A}(W_1) ,\qquad \phi^{\star}(f)= f(\phi) ,$$
and a linear map $\hat{\phi}^{\star} : \Omega (W_2)  \rightarrow   \Omega (W_1)$, 
given by 
 $$ \hat{\phi}^{\star}(\sum^{k_2}_{i=1} f_i(y)dy_i) =  
 \sum^{k_2}_{i=1} \phi^{\star}(f_i)d\phi_i , $$
 where $\phi_i$ is component of $\phi$.
  It is convenient to think of $\hat{\phi}^{\star}$ as a module homomorphism over different rings connected by the homomorphism of rings.   The maps $\phi^{\star}$ and $\hat{\phi}^{\star}$ are usually not distinguished.
 
 We now generalize the classical contact transformations \cite{Ibr}.
 
 \noindent 
{\bf Definition}. Let $U$ be a neighborhood of a point $a\in J^p$ and let $V$ be a neighborhood of a point $b\in J^q$ ($q\leq p$).
An analytic mapping $\phi: U \rightarrow V$ is called a contact mapping if 
the module homomorphism $\hat{\phi}^{\star}$ maps the contact submodule $\mathcal{C}^q_b$ into the contact submodule $\mathcal{C }^p_a$.
%, i.e. $$\hat{\phi}^{\star} (\mathcal{C}^q_b) \subset \mathcal{C}^p_a .$$  
 
The following lifting lemma shows how to construct a contact mapping.
  
 \noindent   
  {\bf Lemma.}  {\it Let $U$ be a neighborhood of the point $a\in J^p$ and 
  let	$\phi: U\rightarrow J^0$ be a analytical mapping of the form
  	$$y =f(x,\dots,u_{\alpha}), \qquad u=g(x,\dots,u_{\alpha}), \qquad |\alpha| \leq p $$
  	such that the matrix $Df=(D_if_j)_{1\leq i,j\leq n}$ is invertible at the point $a$.
  	  	Then there exists an open set $U^1\subset J^{p+1}$ and a unique contact mapping $\phi^1: U^1 \rightarrow J^1$ coinciding with $\phi$ on $ U$.
  	  	}
  
  {\it Proof.} In what follows, we use the following notations 
  $$ dx=(dx_1,\dots,dx_n), \quad du_{\alpha} =(du^1_{\alpha},\dots,du^m_{\alpha}),\quad
  u_{\alpha+1} = (D_i(u^j_{\alpha}))_ {1\leq i\leq n,\ 1\leq j\leq m } .$$
  According to the definition of a contact mapping, the differential form
    $ dv - v_1dx  $ must be represented as
  \begin{equation} \label{dg}
  	dg - v_1df = B_0(du - u_1dx)+\cdots+B_p(du_{\alpha}-u_{\alpha+1}dx) , 
  	\quad |\alpha|=p ,   
  \end{equation}
where   $B_0,...,B_p$ are $m\times m$  matrices.

The left-hand side of equation (\ref{dg}) is written 
	$$g_xdx+g_udu +\cdots + g_{u_{\alpha}}du_{\alpha} - 
	v_1(f_xdx+f_udu +\cdots + f_{u_{\alpha}}du_{\alpha}) , $$
where   $g_x=(\frac{\partial g_j}{\partial x_i}), \ (\frac{\partial g_j}{\partial u^i}), \dots, f_{u_{\alpha}}=(\frac{\partial f_j}{\partial u^i})$ are the corresponding Jacobian matrices. 
We collect together the coefficients of like differential terms in (\ref{dg})
and set all of them equal to zero. The result is a system of matrix equations
 $$ g_x - v_1f_x +B_0u_1+ \dots+B_pu_{\alpha+1} =0 ,$$
$$  g_u -v_1f_u=B_0,\quad \cdots , \quad g_{u_{\alpha}} -v_1f_{u_{\alpha}}=B_p   $$
Substituting $B_0,...,B_p$ into the first equation of this system, we have
 $$ Dg = v_1Df ,      $$
with matrices $Dg =(D_i g_j)$, $Df=(D_i f_k)$ where $j\in \mathbb{N}_m$ and  $i,k\in\mathbb{N}_n$.
 By the condition of our lemma, the matrix $Df$ is invertible. So the lifting formula (first prolongation) is
\begin{equation} \label{v_1}
	v_1 = (Dg)\circ (Df)^{-1}  .
 \end{equation}
The lifting to $J^2,\dots,J^{k+1}$ is carried out in a similar way. The recurrent formula has the form
\begin{equation} \label{v_k}
	v_{k+1} = (Dv_k)\circ (Df)^{-1} . 
\end{equation}
These formulas are  generalizations of the well-known formulas for the lifting \\ (prolongation) of point transformations  \cite{Ovs,Kaptsov}.

\noindent
{\bf  Definition.} If $E=\{f_i\}_{i\in\mathbb{N}_k }$ is a family of differential series of the ring $A^p_a$, then the expression
$$ f_i = 0 , \qquad 1\leq i\leq k $$
is called a system of differential equations and denoted by $sys[E]$.

 \noindent
 {\bf  Definition.} Let $E$ be a family of differential series in the ring $A^p_a$ and let $V$ be an open set in $\mathbb{R}^n$.
  We say that a smooth mapping $s: V \rightarrow J^p$ annihilates the family $E$  if
 \begin{equation} \label{Sol}
 	s^{\star}(E)=0, \qquad \hat{s}^{\star}(\mathcal{C}^p_a)=0
 \end{equation}
 and the rank of  $s$ is $n$ at every point of $V$ .
  If $\pi_0^p$ is the projection of $J^p$ onto $J^0$, then the composition 
  $\pi\circ s$ is called a solution of $sys[E]$.
 
  The mapping $s$ is lifted so that it annihilates the canonical forms (\ref{omega}) as described above. The lifted map is denoted by $\tilde{s}$.
 
 {\bf Proposition.} {\it Let $E_1, E_2$ be two families of differential series of rings $A^p_a$ and $A^p_b$ respectively. Let $U$ be a neighborhood of the point $a_p\in J^p$ and let $V$ be an open set in $\mathbb{R}^n$.
 Assume that a mapping $s: V \rightarrow U$ annihilates  $E_1$ and 
 $\phi: U\rightarrow J^q $ ($q\leq p$) is a contact mapping such that
 $\phi^{\star}(E_2)\subset <E_1>$,  then $\phi\circ s$ annihilates $E_2$. } 

{\it Proof.} Since 
\begin{equation} \label{Ideal}
	\phi^{\star}(E_2)\subset<E_1>, 
\end{equation}
then it is clear that 
$\tilde{s}^{\star}(\phi^{\star}(E_2))= 0$.
It follows that
$$ \tilde{s}^{\star}(E_2(\phi))= E_2(\phi\circ s)= (\phi\circ s)^{\star}(E_2)=0 .   $$
The equality 
$(\phi\circ s)^{\star}(\mathcal{C}^p_b)=0 $
follows in the same way. 

\noindent
{\it Remarks.} To put it simply, the contact mapping $\phi$ maps solutions of $sys[E_1]$ to solutions of $sys[E_2]$ if $\phi^{\star}(E_2)\subset<E_1>$.
If we extend the homomorphism $\phi^{\star}$ to the ideal $<E_2>$, then the condition (\ref{Ideal}) can be done more invariant 
$$\tilde{\phi^{\star}}(<E_2>) \subset<E_1> .$$

{\bf Definition.} {\it Let $E$ be a family of differential series of the ring $A^p_a$ and let $U$ be a neighborhood of the point $a_p\in J^p$.
	A contact mapping $\phi: U\rightarrow J^q$ ($q\leq p$) such that $\phi^{\star}(E)\subset <E>$ is called a symmetry of $sys[E]$ . }

Let us give examples of contact mappings connecting partial differential equations. We now use the classical notation.
Consider two equations
\begin{equation} \label{u_tt}
	u_{tt} = x^n u_{xx}  ,\qquad n\in\mathbb{N},
\end{equation}
\begin{equation} \label{v_tt}
	v_{tt} = v_{yy} +\frac{m}{y}v_y  ,\qquad m\in\mathbb{R} .
\end{equation}
We want to find a contact mapping that transforms solutions of the equation (\ref{u_tt}) to solutions of the equation (\ref{v_tt}).
Consider a mapping $\phi: J^1\rightarrow J^0$ of the form
\begin{equation} \label{phi}
	t^{\prime} = t, \qquad y= h(x), \qquad v=f(x)u_x +g(x)u ,
\end{equation}
where $h, f, g$ are some smooth functions.
We will lift this mapping according to the formulas (\ref{v_1}), (\ref{v_k})
$$  v_t = D_t v = fu_{tx}+gu_t , \qquad  v_{tt} = fu_{ttx}+gu_{tt} ,$$
\begin{equation} \label{v_yy}
	v_y = \frac{D_x(v)}{D_x h}= 
\frac{D_x(fu_x+gu)}{ h^{\prime}} ,\qquad v_{yy} = \frac{D_x(v_y)}{h^{\prime}} . 
\end{equation}
Substituting the found expression for $v_{tt}$ into (\ref{v_tt}), we have
$$ fu_{ttx}+gu_{tt} = v_{yy} +\frac{m}{y}v_y .$$
We can express  $u_{tt}, u_{ttx}$ by using (\ref{v_yy}) and obtain a new equation
\begin{equation} \label{v_xxx}
	(x^nu_{xx})_xf + x^nu_{xx}g+ \frac{1}{h^{\prime}}D_x\left(\frac{D_x(fu_x+gu)}{h^{\prime}} \right) +
	\frac{mD_x(fu_x+gu)}{hh^{\prime}} =0 .
\end{equation}
The left-hand side of this equation is a polynomial in $u_{xxx}, u_{xx}, u_x, u$.
Collecting the coefficients of like terms in the polynomial and setting all of them equal to zero, we obtain four equations for the functions $f, h, g$. 
The two shortest equations are
$$  x^n (h^{\prime})^2 = 1 , \qquad m (h^{\prime})^2 g^{\prime} +hh^{\prime}g^{\prime\prime} -hh^{\prime\prime}g^{\prime}  = 0 .        $$
Integrating these equations for $n\neq 2$, we find
 $$ h= \pm \frac{2}{2-n} x^{\frac{2-n}{2}} + c_0 , \qquad g= c_1+c_2h^{ 1-m}$$
where $c_0, c_1, c_2$ are arbitrary constants.
The remaining two equations for the function $f$ are easy to integrate.
The following two cases  arise:
 $c_1\neq 0, c_2=0$ and $c_1= 0, c_2 \neq 0$.
 In the first case, the function $f$ is equal to $ax$ $(a\in\mathbb{R})$.
Then the transformation 
 $$y= \pm \frac{2}{2-n} x^{\frac{2-n}{2}} ,\qquad v=a(xu_x +(n-1)u)  , a\in\mathbb{R}   $$
 maps solutions of the equation  (\ref{u_tt}) into ones of equation 
 $$ v_{tt}=v_{yy}+\frac{3n-4}{2-n}v_y .$$
 
 In the second case, the  transformation 
 $$ y= \pm \frac{2}{2-n} x^{\frac{2-n}{2}} ,\qquad  v= ax^{2n-3}(xu_x+(n-1)u)    $$
 maps solutions of the equation  (\ref{u_tt}) into ones of equation 
 $$ v_{tt}=v_{yy} +\frac{5n-8}{n-2}v_y . $$

 \section{ Parametric contact mappings}
 
 It is well known that 
 finding symmetries of differential equations can be simplified if we restrict ourselves to the search for one-parameter groups of transformations that leave the equations invariant.
  In this section, it is assumed that contact mappings depend on the parameter $a$. More precisely, we seek an expansion of the mappings in powers of $a$.
 
Next we restrict ourselves to to the case $n=2, m=1$ and use the classical notation for coordinates in the jet spaces $J^0(x,y,u)$, $J^1(x,y,u,p,q)$, \\ $J^2(x,y,u, p,q,r,s,t)$.
 
Consider a mapping of the form 
 \begin{align}
 	\bar{x}= x +ax_1+a^2x_2+a^3x_3+ \dots  \ ,\notag  \\ 
 	\bar{y}= y + ay_1+a^2y_2+a^3y_3+ \dots \ , \label{Y=} \\ 
 	\bar{u}= u + au_1+a^2u_2+a^3u_3+ \dots \ ,  \notag
 \end{align}
where $x_1, x_2, x_3,y_1,\dots,u_3$ are functions of $x, y,..., u_{\alpha}$.
To find the first prolongation of the mapping (\ref{Y=})
$$\bar{p}= p + ap_1+a^2p_2+a^3p_3+ \dots \ , $$
$$\bar{q}= q + aq_1+a^2q_2+a^3q_3+ \dots \ , $$
it is necessary that the differential form
\begin{equation} \label{w_0}
	\bar{\omega}_0= d\bar{u}-\bar{p}d\bar{x}-\bar{q}d\bar{y}
\end{equation}
vanishes when the Pfaff equation
 \begin{equation} \label{w_0}
	\omega_0=du-pdx-qdy =0  
\end{equation}
 is satisfied.
 
Substituting the expressions (\ref{Y=}) into the form $\bar{\omega}_0$, by using the equality (\ref{w_0}), and collecting  together all
terms that contain $a$, we obtain the well-known the first prolongation formulas \cite{Ovs}
$$ p_1 = D_x(u_1) - pD_x(x_1)-qD_x(y_1)  , \quad
q_1 =  D_y(u_1) - pD_y(x_1)-qD_y(y_1)  .   $$
Collecting  together all terms that contain $a^2$, we find that
\begin{gather*} 
	p_2 =D_x(u_2) - pD_x(x_2)- p_1D_x(x_1)-qD_x(y_2) -q_1D_x(y_1) ,\\
	q_2 = D_y(u_2) - pD_y(x_2)- p_1D_y(x_1)-qD_y(y_2) -q_1D_y(y_1) . 
\end{gather*} 
It is important to remark that  $x_2,y_2,u_2$ are an arbitrary
functions.
When we collect together all terms that contain $a^3$ this leads to
\begin{gather*} p_3 =D_x(u_3) - pD_x(x_3)- p_1D_x(x_2)- p_2D_x(x_1)-qD_x(y_3) -q1D_x(y_2) -q_2D_x(y_1),\\
	q_3 =D_y(u_3) - pD_y(x_3)- p_1D_y(x_2)- p_2D_y(x_1)-qD_y(y_3) -q_1D_y(y_2) -q_2D_y(y_1)  . 
\end{gather*}
Similar formulas are valid for $p_n, q_n$ $(n>3)$.

It is easy to find formulas for the second prolongation
\begin{gather*}   \bar{r}= r + ar_1+a^2r_2+a^3r_3+ \dots \ ,\qquad   
	\bar{s}= s + as_1+a^2s_2+a^3s_3+ \dots \ , \\
	\bar{t}= t + at_1+a^2t_2+a^3t_3+ \dots \ .  \end{gather*} 
For this to be accomplished, it is necessary that the differential forms
$$ \bar{\omega}_{10}= d\bar{p} -\bar{r}d\bar{x} - \bar{s}d\bar{y} \qquad 
\bar{\omega}_{01}=  d\bar{q} -\bar{s}d\bar{x} - \bar{t}d\bar{y}  $$
vanish if  
$$   \omega_{10}= dp -rdx - sdy = 0  , \qquad  \omega_{01}=  dq -sdx - tdy = 0 . $$
Using arguments similar to those given above, it is easy to obtain the following formulas
\begin{gather*}
	r_1 = D_x(p_1)-rD_x(x_1)-sD_x(y_1) , \quad s_1=D_y(p_1)-rD_y(x_1)-sD_y(y_1) ,\\
	t_1=D_y(q_1)-sD_y(x_1)-tD_y(y_1) , \\
	r_2 = D_x(p_2) -rD_x(x_2)-r_1D_x(x_1) - sD_x(y_2) -s_1D_x(y_1) ,\\ 
	s_2 = D_y(p_2) -rD_y(x_2)-r_1D_y(x_1) - sD_y(y_2) -s_1D_y(y_1) , \\ 
	t_2=  D_y(q_2) - sD_y(x_2) -s_1D_y(x_1) -tD_y(y_2) - t_1D_y(y_1) . 
\end{gather*} 

As example, consider the Burgers equation
\begin{equation} \label{Burg}
	u_y - u_{xx} - uu_x =0 .
\end{equation}
We look for contact mappings such that (\ref{Burg}) is invariant under the 
ones. 
The symmetry condition implies that the expression
$$ \bar{u}_{\bar{y}} - \bar{u}_{\bar{xx}} -\bar{u}\bar{u}_{\bar{x}} $$
lies in the ideal $<u_t - u_{xx}-uu_{x} > .$
%This means that the equation
% \begin{equation} \label{q bar}
%\bar{u}_{\bar{y}} - \bar{u}_{\bar{xx}} -\bar{u}\bar{u}_{\bar{x}} =0 
%\end{equation}
%follows from (\ref{Burg}). 

The simplest of these mappings has the form
$$ \bar{x}= x,\qquad \bar{y}=y,\qquad \bar{u}= u+ \frac{2au_x}{au+1} , \quad a\in\mathbb{R} .$$
This mapping satisfies the second-order differential equation
$$ \bar{u}_{aa}= 2 \frac{ \bar{u}_{a}(a \bar{u}_{a}-\bar{u})}{a\bar{u }+1} $$
with initial conditions: $\bar{u}(0)=u, \bar{u}_a(0)=u_x $.
Recall that in Lie theory, symmetry transformations satisfy  first order ordinary differential equations \cite{Ovs}.

A more general symmetry mapping is given by the formulas
$$  \bar{x}= x,\qquad \bar{y}=y,\qquad \bar{u}= u+ 2D_x(\log h) ,        $$
where the function $h$ satisfies the condition 
\begin{equation}  \label{h}
	D_y h - D_x^2h -u D_xh\in <u_y-u_{xx}-uu_x>.    
\end{equation}
More precisely, the following statement is true.

\noindent
{\bf Proposition.} {\it Let $u$ be a solution to the equation (\ref{Burg}), and let the differential series $h$ satisfy the condition 	(\ref{h}).
Then the function
\begin{equation} \label{v}
	v = u +2D_x(\log h)
\end{equation}
is also a solution to the Burgers equation
$$ v_t-v_{xx}-vv_x=0. $$ }	
Indeed, substituting the function $v$ given by (\ref{v}) into the left-hand side of the last equation, we obtain an expression that can be represented as
$$ u_y - u_{xx}-uu_x + 2D_x\left(\frac{D_yh -D_{xx}h -uD_xh}{h} \right) .$$	
Thus the Proposition follows from (\ref{h}).

It is important to note that if $h$ satisfies the condition (\ref{h}), then $\eta=D_x h$ is a solution of the determining equations for 
 the  symmetry generator. Therefore, knowing the symmetries of the equation
it is easy to find $h$.
 
 In particular, the condition (\ref{h}) is satisfied by $h$ of the form
 \begin{gather*}
 	h =  s_0 + a[s_1(2p+u^2) +s_2u +s_3(yu+x) +s_4(2yp+yu^2+xu) +\\ 
 	s_5(y^2(4p+2u^2) +2xyu +x^2 +2y ) ] ,
 \end{gather*}
 where $a, s_0,...,s_5$ are arbitrary constants. 
 If $s_0\neq 0$, then the function $\bar{u}$ is represented by
 power series in $a$.

 This work is supported by the Krasnoyarsk Mathematical Center and financed by the Ministry of Science and Higher Education of the Russian Federation in the framework of the establishment and development of regional Centers for Mathematics Research and Education (Agreement No. 075-02-2023-912).

\end{document}